\documentclass[
reprint,
superscriptaddress,
 aps,
]{revtex4-1}
\usepackage{graphicx}
\usepackage{bm}
\usepackage{xr-hyper}
\usepackage{hyperref}
\usepackage{siunitx}
\usepackage{physics}

\begin{document}

\preprint{APS/123-QED}

\title{Inertial spin dynamics in epitaxial cobalt films}

\author{Vivek Unikandanunni}
\affiliation{Department of Physics, Stockholm University, 10691 Stockholm, Sweden}
\author{Rajasekhar Medapalli}
\affiliation{Center for Memory and Recording Research, University of California San Diego, San Diego, CA 92093, USA}
\affiliation{Department of Physics, Lancaster University, Bailrigg, Lancaster LA1 4YW, United Kingdom}
\author{Marco Asa}
\affiliation{Department of Physics, Politecnico di Milano Technical University, Milano, Italy}
\author{Edoardo Albisetti}
\affiliation{Department of Physics, Politecnico di Milano Technical University, Milano, Italy}
\author{Daniela Petti}
\affiliation{Department of Physics, Politecnico di Milano Technical University, Milano, Italy}
\author{Riccardo Bertacco}
\affiliation{Department of Physics, Politecnico di Milano Technical University, Milano, Italy}
\author{Eric E. Fullerton}
\affiliation{Center for Memory and Recording Research, University of California San Diego, San Diego, CA 92093, USA}
\author{Stefano Bonetti}
\email{stefano.bonetti@fysik.su.se}
\affiliation{Department of Physics, Stockholm University, 10691 Stockholm, Sweden}
\affiliation{Department of Molecular Sciences and Nanosystems, Ca' Foscari University of Venice, 30172 Venice, Italy}

\begin{abstract}
We investigate the spin dynamics driven by terahertz magnetic fields in epitaxial thin films of cobalt in its three crystalline phases. The terahertz magnetic field generates a torque on the magnetization which causes it to precess for about 1 ps, with a sub-picosecond temporal lag from the driving force. Then, the magnetization undergoes natural damped THz oscillations at a frequency characteristic of the crystalline phase. We describe the experimental observations solving the inertial Landau-Lifshitz-Gilbert equation. Using the results from the relativistic theory of magnetic inertia, we find that the angular momentum relaxation time $\eta$ is the only material parameter needed to describe all the experimental evidence. Our experiments suggest a proportionality between $\eta$ and the strength of the magneto-crystalline anisotropy.

\end{abstract}

\maketitle

The fundamental understanding of magnetism has much improved since the first experiments on magnetic materials at femto- and picosecond time scales, the so-called ultrafast regime. The pioneering work of Beaurepaire $\textit{et al.}$ \cite{beaurepaire1996ultrafast} demonstrated that sub-picosecond magnetization dynamics is possible, against the prediction of the textbook Landau-Lifshitz-Gilbert (LLG) equation. Those results have since then been confirmed in several experiments and using both optical \cite{koopmans2010explaining,kirilyuk2010ultrafast,carva2013ab,bigot2009coherent,kim2012ultrafast,eschenlohr2013ultrafast,bigot2005ultrafast,koopmans2000ultrafast,hohlfeld1997nonequilibrium,cinchetti2006spin,guidoni2002magneto} and X-ray techniques \cite{stamm2007femtosecond,boeglin2010distinguishing,graves2013nanoscale,iacocca2019spin}, making the field of ultrafast magnetism, and of magnetization dynamics beyond the LLG equation, an active field of research with possible implications for novel data storage technologies.

Another recent analysis of the LLG equation showed that from a classical mechanics point of view, the LLG equation assumes an unphysical inertial tensor \cite{ciornei2011magnetization}, also suggested by Gilbert himself in a footnote almost two decades ago \cite{gilbert2004phenomenological}. However, Ciornei \emph{et al.} \cite{ciornei2011magnetization} re-derived the equation with a realistic inertial tensor which resulted in a slightly revised equation known as the inertial LLG (iLLG) equation, which predicts the appearance of a nutation resonance at a frequency much higher than the ferromagnetic resonance (FMR) one. After that pioneering work, much theoretical effort has been performed trying to identify the frequency regime where such nutation resonance was to be expected, with predictions varying over a few orders of magnitude \cite{fahnle2011generalized,bottcher2012significance,li2015inertial,kikuchi2015spin,bastardis2018magnetization}. The experimental confirmation has been achieved only recently and detected nutation dynamics in the $\sim1$ THz range \cite{neeraj2021inertial}. Despite this novel experimental evidence, and the significant theoretical progress to understand the microscopic origin of inertia \cite{bhattacharjee2012atomistic,olive2012beyond,olive2015deviation,fransson2017microscopic,mondal2017relativistic,cherkasskii2020nutation,cherkasskii2021dispersion,thibaudeau2021emerging,mondal2021influence,mondal2021theroy,mondal2021nutation,lomonosov2021anatomy,anders2020versatile,ruggeri2021numerical,titov2021deterministic,titov2021inertial,giordano2020derivation,rahman2021observable,gupta2021co2feal,jhuria2020spin}, a complete picture of how different material parameters affect the nutation dynamics is still missing.

In this Letter, we provide the first experimental data on the dependence of inertial spin dynamics on a key magnetic property, i.e. the magneto-crystalline anisotropy. It has been suggested that spin-orbit coupling is the fundamental interaction needed to derive a correct inertial tensor from first principles. Hence, experiments where the magneto-crystalline anisotropy is well defined and controlled, are expected to return important insights on this open question. Similar to Ref.~\cite{neeraj2021inertial}, we perform THz pump / optical probe time-resolved magneto-optical measurements to trigger and detect inertial spin dynamics. In contrast to the approach of Ref.~\cite{neeraj2021inertial}, where a narrowband ($\Delta f/f\approx0.1$) THz source was implemented, we use intense single-cycle terahertz radiation \cite{hoffmann2011intense} whose broad band ($\Delta f/f>1$) allows to cover the frequency range where the nutation resonance is expected to appear. In addition, due to the impulsive character of the driving pulse, our measurements are expected to detect not only the forced response of the system, but also its natural one. We model the magnetization dynamics solving the inertial LLG equation numerically and we contextualize our results with the existing microscopic theory of magnetic inertia.

\begin{figure}[t]
\includegraphics[width=\columnwidth]{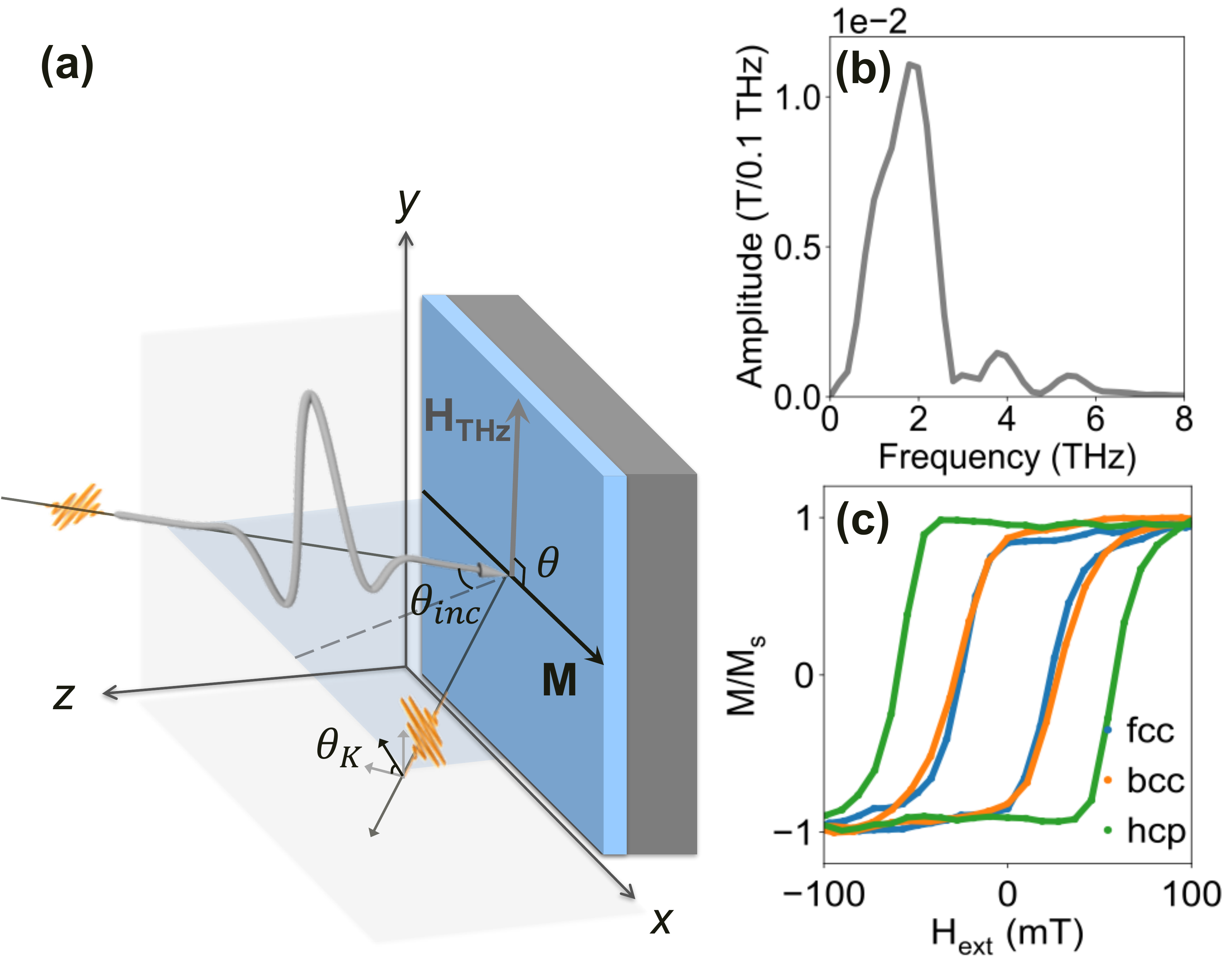}
\caption{\label{fig:fig1}(a) Geometry of THz pump-MOKE probe setup. (b) Frequency spectrum of terahertz pump pulse (c) Magnetization loops for fcc, bcc, and hcp cobalt measured using the longitudinal MOKE.}
\end{figure}

We choose to investigate three epitaxial cobalt thin films grown on MgO substrate with face-centered cubic (fcc), body-centered cubic (bcc), and hexagonal close packed (hcp) crystal structures. The fabrication details for the three samples are given in the Supplemental Material \cite{SM}. The hcp sample has a strong in-plane magneto-crystalline anisotropy characterized by an easy magnetization axis along the $c$-direction of the hcp structure (which lies in the film plane in our films), and a hard axis orthogonal to it. For the two cubic crystal structures, the anisotropy is still in-plane, but its strength is much reduced, and a hard magnetization direction is not clearly identified \cite{SM}. The hcp and fcc samples were grown under similar deposition conditions and are respectively 10 nm and 15 nm in thickness, while the bcc sample was grown in a different laboratory and it has a thickness of 8 nm. Fig.~\ref{fig:fig1}(a) shows the geometry of the single cycle THz pump - optical probe experiment. The magnetization $\mathbf{M}$ of the sample is aligned along the $x$-direction by means of an external bias field $|H_{ext}|=100$ mT, kept constant during the experiment. The single-cycle THz pump pulses are generated in the organic crystal OH1 by the optical rectification of 1300 nm radiation from an optical parametric amplifier \cite{jazbinsek2019organic}. The pump pulse has a peak magnetic field of 0.3 T parallel to the $y$-direction, which maximizes the torque on the magnetization, and impinges on the sample at an angle of incidence $\theta_{inc}=45$ degrees. Fig.~\ref{fig:fig1}(b) is the Fourier transform of the electro-optical sampling measurement in a a 50 $\mu$m-thick GaP crystal \cite{nahata1996coherent}, used to characterize the THz pulse. It shows that the pump field is peaked at around 2 THz, and has a bandwidth exceeding 1 THz. The magnetization dynamics is probed using the time-resolved magneto-optical Kerr effect (MOKE), in the specific measuring the Kerr rotation angle $\theta_K$ of a nominally 40 fs, 800 nm probe beam, using a balanced detection scheme. All radiation is derived from the same amplified laser system ensuring intrinsic synchronization, with the relative delay between the beams controlled by a mechanical translation stage, and with the pump modulated at a frequency equal to half the laser repetition rate. Fig.~\ref{fig:fig1}(c) shows the easy axis magnetization loops for the three samples investigated in this work. The coercive field for the hcp sample is about 50 mT, whereas is approximately 30 mT for both fcc and bcc samples \cite{SM}.

\begin{figure}[t]
\includegraphics[width=\columnwidth]{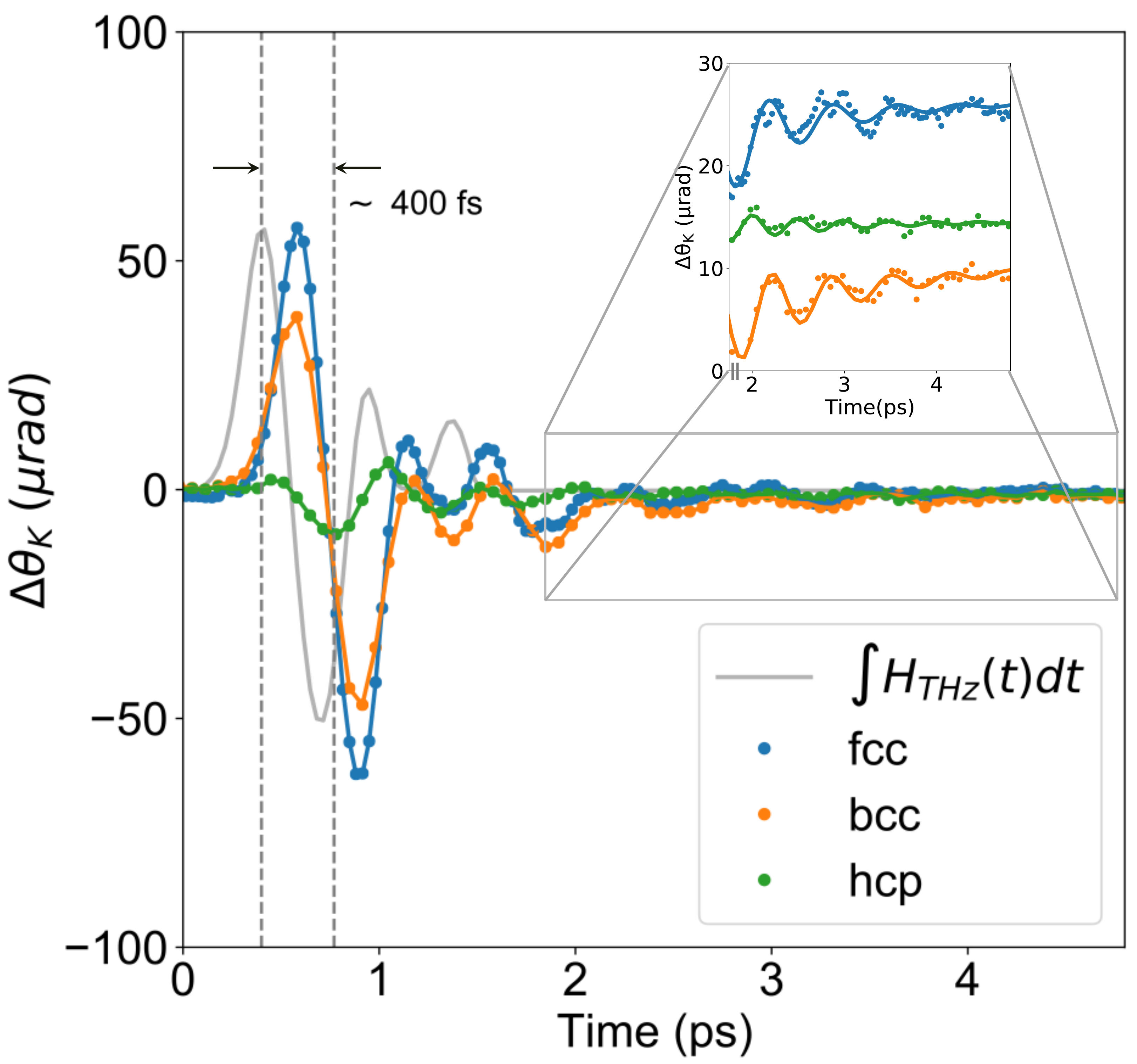}
\caption{\label{fig:fig2}Solid symbols: time-resolved Kerr rotation measurements on fcc, bcc and hcp cobalt thin films. Dashed line: integral of the pump THz magnetic field $H_{THz}$. Inset: zoomed-in main panel data for $t>1.7$ ps. The data is shifted vertically for clarity. The continuous lines the best fits obtained using Eq.~\eqref{eq2}.}
\end{figure}

Fig.~\ref{fig:fig2} shows the time-resolved MOKE measurements of the terahertz-field induced dynamics in all three samples. The plotted traces represent the difference of the data recorded with magnetic fields of equal magnitude but opposite polarity, ensuring the magnetic character of the signal \cite{hudl2019nonlinear}. For all samples, the MOKE response is dominated by the coherent precession of the magnetization around the applied THz magnetic field $H_{\rm THz}$, which in fcc and bcc films is larger in amplitude than in the hcp one. A very small demagnetization, showing up as a lingering non-zero average MOKE signal, is also present. The presence of both coherent (precession) and incoherent (demagnetization) effects in the observed THz-driven dynamics is consistent with Ref.~\cite{bonetti2016thz}, where it was also observed that in epitaxial films the demagnetization signal was negligible. Similar to Refs. \cite{bonetti2016thz} and \cite{neeraj2021inertial}, no coherent precession is observed when $H_{\rm THz}\parallel\mathbf{M}$, since in that case the torque acting on the magnetization is zero. We present this measurement in the Supplemental Material \cite{SM}.

The dashed grey line in the same plot is the integral of $H_{\rm THz}$ over time, obtained numerically from the electro-optic sampling measurement used to characterize the THz field. Ref. \cite{bonetti2016thz} demonstrated that the coherent response of the magnetization to an off-resonant THz field can be obtained simply by integrating $H_{\rm THz}$, which is the solution of the LLG equation for small and off-resonant excitations. However, while in Ref. \cite{bonetti2016thz} the temporal overlap between the MOKE data and the integral of $H_{\rm THz}$ was exact within the experimental error, here we notice a substantial lag between them, approximately 200 fs for the fcc and bcc samples, and 400 fs for the hcp one, highlighted by the vertical lines. In other words, the shape of the MOKE response is still consistent with the integral of the THz field if properly scaled, however its phase is not. This phase shift is particularly dramatic for the hcp sample, where it looks as if the magnetization precesses in the opposite direction as compared to the fcc and bcc samples. We repeated the experiment on all samples, and the evidence is robust. The sign of the magneto-optical coefficient does not change either between the different samples, as demonstrated by the magneto-optical hysteresis loops in Fig. \ref{fig:fig1}(c). Hence, the observed phase shift is real and it appears to be strongly dependent on the crystalline structure of the sample.

Another intriguing observation from the data in Fig.~\ref{fig:fig2} is found in the inset, where we zoom in on the main panel data at temporal delays $t>1.7$ ps. When the pump field has left the sample, a comparatively tiny, yet detectable, damped ringing of the magnetization can be observed. We can fit such behavior with the phenomenological formula
\begin{equation}\label{eq2}
\Delta \theta_K(t) = Ae^{-t/\tau_1}+Be^{-t/\tau_2}\sin(2\pi ft)
\end{equation}
where $\tau_1$ is the recovery time of the incoherent demagnetization dynamics, $\tau_2$ is the decay time of the sinusoidal oscillation, $f$ is the frequency of the oscillations, and $A$, $B$ are the constants describing the the amplitude of the demagnetization and, respectively, of the sinusoidal oscillations. The fit returns $f_{fcc}\approx1.3$ THz, $f_{bcc}\approx1.4$ THz and $f_{hcp}\approx2.1$ THz. $\tau_2$ is found to be approximately 0.82, 0.70 and 0.72 ps for the fcc, bcc and, respectively, hcp samples, corresponding to damping coefficients $\alpha=1/\omega\tau_2$ which are $\alpha_{fcc}\approx0.15$, $\alpha_{bcc}\approx0.16$ and $\alpha_{hcp}\approx0.10$.

\begin{figure}[t]
\includegraphics[width=\columnwidth]{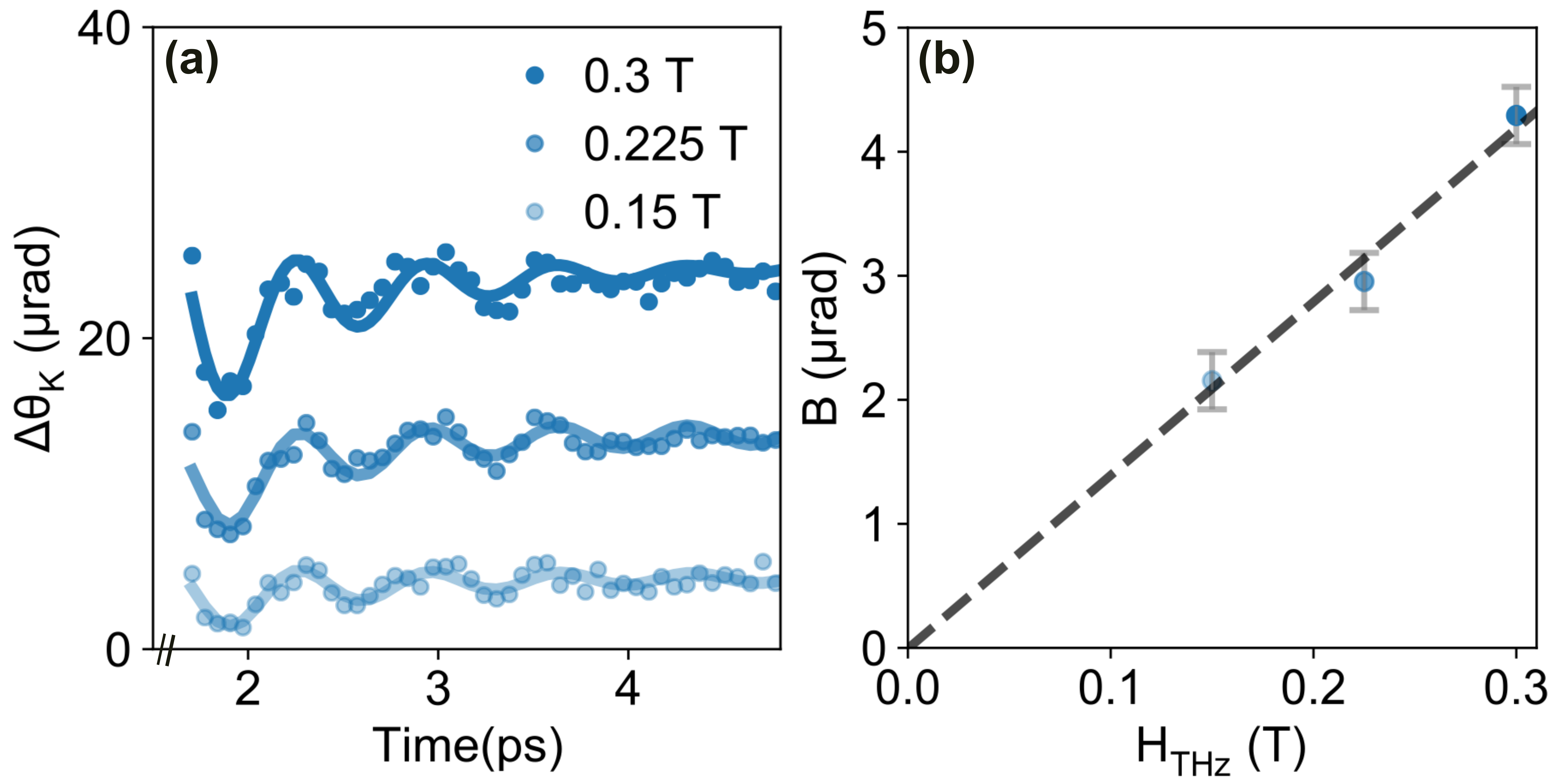}
\caption{\label{fig:fig3} (a) Symbols: time-resolved Kerr signal at $t>1.7$ ps for THz magnetic field values of different maximum amplitude. The data is vertically shifted for clarity. Solid lines: best fit obtained using Eq. \eqref{eq2}. (b) Symbols: extracted oscillation amplitude $B$ as a function of THz magnetic field and corresponding standard deviation. Dashed line: linear fit to the data with imposed zero offset.}
\end{figure}

Before discussing these results, we present in Fig.~\ref{fig:fig3} the THz field dependent measurement on the fcc sample, which showed the largest signal in Fig.~\ref{fig:fig2}. The terahertz field strength is controlled through the relative orientation of a pair of wire-grid polarizers in the THz pump path. The second polarizer was kept fixed in order to preserve the polarization of the THz field impinging on the sample. The time-resolved MOKE signal is shown in Fig. \ref{fig:fig3}(a) for the maximum field strength, 75\% and 50\% of it, below which we were at the noise level of our setup. We used again Eq.~\eqref{eq2} to fit the oscillations and to extract the amplitudes and recovery times as a function of THz field strength. Fig.~\ref{fig:fig3}(b) shows the extracted oscillations amplitude $B$ as a function of THz field strength, which can be fitted with a linear function with no offset.

The evidence presented so far is consistent with the presence of a sizeable magnetic inertia in crystalline cobalt films, manifesting itself with a lagging response to an external field and to the appearance of nutation oscillations. In order to investigate this hypothesis thoroughly, we performed numerical simulations using the inertial LLG equation, written in a slightly different form than the one given in Ref. \cite{neeraj2021inertial}
\begin{equation}\label{eq3}
\frac{d\mathbf{M}}{dt} = -\abs{\gamma}\mathbf{M}\times\mathbf{H}_{\rm eff}+\mathbf{M}\times\left(\alpha\frac{d\mathbf{M}}{dt}-\eta\frac{d^2\mathbf{M}}{dt^2}\right),
\end{equation}
where $\abs{\gamma}$/$2\pi$ = 28 GHz/T is the gyromagnetic ratio, $\mathbf{H}_{\rm eff}=(H_{\rm bias}+H_K)\mathbf{x}+H_{\rm THz}(t)\mathbf{y}+H_{d}\mathbf{z}$ is the effective magnetic field which comprises of the external bias field $H_{\rm bias}$, the anisotropy field $H_K$, the applied THz field $H_{\rm THz}(t)$, and the demagnetizing field $H_{d}$; $M_s$ is the saturation magnetization of the sample, $\alpha$ is the Gilbert damping parameter, and $\eta$ is the angular momentum relaxation time defined as in Ref. \cite{mondal2021nutation}, i.e. $\eta=\alpha\tau$. Since $\alpha\ll1$, the absolute values $\eta$ are much smaller than $\tau$ reported in Ref. \cite{neeraj2021inertial}. The last term on the right hand side of Eq.~\eqref{eq3} is the nutation term that is present only when $\eta\neq0$. In the following, we solve this equation in the macrospin approximation (i.e. the sample is considered as a homogeneous ferromagnet) and using a conventional fourth-order Runge-Kutta method.

\begin{figure}[t]
\includegraphics[width=\columnwidth]{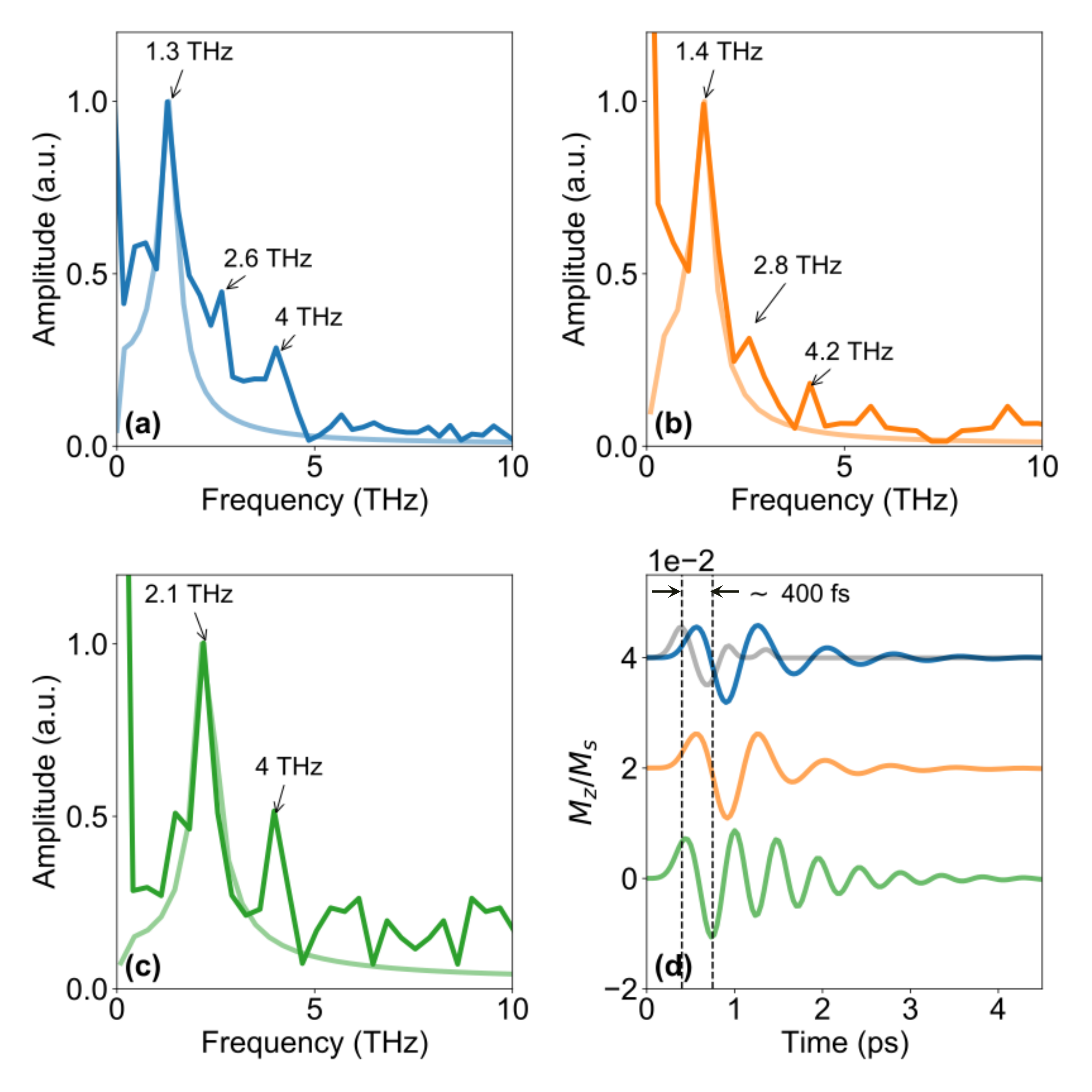}
\caption{\label{fig:fig4} Experimental (solid) and simulated (semi-transparent) Fourier transform of the magnetization dynamics in (a) fcc (blue) (b) bcc (orange) and (c) hcp (green) cobalt thin films. The full experimental trace is used for the experimental data in Fig. \ref{fig:fig2}, and the $M_z$ component for the simulations. (d) Solid lines: simulated response of $M_z$ to a single cycle terahertz field $H_{\rm THz}$ in the time-domain using the same parameters. Dashed line: integral over time of $H_{\rm THz}$.}
\end{figure}

In Fig.~\ref{fig:fig4}(a)-(c), we compare the results from these numerical simulations to the experimental results in frequency domain. For both simulations and experiments, we Fourier transform the temporal traces obtained at time delays $t>1.7$ ps, when the THz pump field has left the sample. Including the full temporal trace would hide the small features below the broad single-cycle response. In the numerical simulations, we calculate $\mathbf{H}_{\rm eff}$ solely from experimentally measured quantities found in the Supplemental Material \cite{SM} or in previous references \cite{unikandanunni2021anisotropic}. This allow us to estimate $H_d\approx1.6$ T for all samples, and $H_K\approx0.8$ T for the hcp sample and one order of magnitude smaller for the other two samples. We used the nominal values for $H_{\rm bias}$ and $H_{\rm THz}$. The only two free parameters are then $\eta$ and $\alpha$, which can be independently tuned to match the peak frequency and, respectively, linewidth. Using $\eta_{fcc}=120$ fs, $\eta_{bcc}=110$ fs and $\eta_{hcp}=75$ fs we can reproduce the main experimental peak frequency, and assuming $\alpha_{fcc}=0.15$, $\alpha_{bcc}=0.16$ and $\alpha_{hcp}=0.10$ from the fits using Eq.~\eqref{eq2}, we can also match the linewidth of the main peak. No observable difference was found within 5-10 fs for $\eta$ and within 0.01 for $\alpha$, giving an approximate 10\% relative accuracy. We have performed additional simulations (not shown) and we also observe that the effective field does not affect the nutation frequency and linewidth in a noticeable way unless it reaches values of the order of a few Tesla.

We discuss below the plausibility of these values; assuming for the time being that they are reasonable, and looking at Fig.~\ref{fig:fig4}(d), we obtain the remarkable result that the inertial LLG equation is able to reproduce all the experimental evidence of Fig.~\ref{fig:fig2}: the presence of a damped nutation oscillation and the temporal shift of the coherent magnetization precession. In this small amplitude limit, the inertial LLG also predicts a linear scaling of the coherent precession and of the nutation amplitude with terahertz field strength, as shown experimentally in Fig.~\ref{fig:fig3}(b). None of these experimental evidences can be reproduced solving the standard LLG equation, proving that the additional inertial term is necessary.

The only experimental evidence which is not reproduced by the inertial LLG equation, in the currently known form and in the macrospin approximation, is the presence of higher order harmonic peaks in the frequency response seen in Fig.~\ref{fig:fig4}(a)-(c). We can clearly identify the second and third harmonics for the fcc and bcc samples, and the second harmonic for the hcp one. However, Kikuchi $\textit{et al.}$ \cite{kikuchi2015spin} predicted such possibility if the third and other higher order time derivatives of the magnetization, not included in the standard framework of the inertial LLG model, are considered. 
We leave this question open to future theoretical and experimental investigations, here we simply note that the presence of harmonics at integer multiples $n$ of the fundamental frequency could be consistent with nutation dynamics. It is not consistent with the presence of standing waves across the film thickness, which show instead a $n^2$ dependence due to confinement \cite{razdolski2017nanoscale}. We also do not observe any apparent inverse thickness dependence, which is instead expected in the case of standing waves.

As a final control to test the general validity of our experimental results and of the inertial LLG equation, we performed additional measurements using a different THz single-cycle pump field with a bandwidth extending from 2 to 4 THz instead, i.e. with negligible overlap with the nutation resonances. This is achieved by replacing the nonlinear crystal generating the THz radiation and by adjusting the corresponding pump wavelength, leaving the rest of the setup unchanged. The results are reported in the Supplemental Material \cite{SM}, and they show that neither THz oscillations nor phase shift of the coherent precession is observed when the pump field does not match the nutation resonance. This is also in agreement with previous measurements done in fcc cobalt driven by a THz field with similar bandwidth \cite{shalaby2018coherent}. The inertial LLG equation with the same parameters reproduces even these experimental data to an excellent degree, with no nutation oscillations nor phase shift observed in this case.

We now turn the discussion to the two free parameters in the inertial LLG equation, namely the damping $\alpha$ and the angular relaxation time $\eta$. In order to match the experimental linewidth, we used for all three films a damping parameter which is an order of magnitude larger than the typical FMR Gilbert damping of these materials \cite{unikandanunni2021anisotropic}. While we do not have a microscopic explanation for these large values, we notice that the same issue was found in the first experimental report of nutation in ferromagnets and left as an open question \cite{neeraj2021inertial}. Our experiments, which are able to observe the natural nutation oscillations, allow us to extract the damping factor directly from the data. In order to estimate the magnitude of $\alpha$ for the inertial dynamics, a microscopic theoretical investigation is needed, which is beyond the scope of our work. Our experiments suggest though that a complete inertial LLG equation may contain either distinct Gilbert and inertial damping coefficients, or a time-dependent one, in order to fully describe the magnetization dynamics. A time-dependent $\alpha$ can be qualitatively linked to a damping mechanism dominated by comparatively stronger electron-phonon scattering at sub-picosecond time scales, and weaker spin-lattice relaxation at longer time scales \cite{suhl1998theory,gilbert2004phenomenological,ritzmann2020theory}. A time-dependent $\alpha$ has also been recently suggested to include all damping mechanisms, including time-retardation effects \cite{bajpai2019time}.

The most important experimental observation of this work, which is expected to contribute to a microscopic understanding of inertial dynamics, is the strong dependence of the nutation frequency on the different cobalt samples with different magneto-crystalline anisotropy, which in turn is dependent on the strength of the spin-orbit coupling. The relativistic theory of magnetic inertia at ultrafast time scales \cite{mondal2017relativistic} demonstrated that the presence of a finite angular momentum relaxation time is due to a spin-orbit coupling effect of order $1/c^4$, whereas the Gilbert damping is of order $1/c^2$. These two quantities are therefore dependent on each other, and it was suggested in Ref.~\cite{mondal2017relativistic} that the ratio $\eta/\alpha$ should be a constant. From our data, we calculate $\eta/\alpha=746\pm46$ fs from the three cobalt films, which is constant within the accuracy of our estimates of the two parameters ($\sim10\%$). This reduces the number of free parameters in the inertial LLG equation to only one, at least within the same $3d$ element, and it further strengthens the interpretation of our results in terms of relativistic spin dynamics. We note that the magneto-crystalline anisotropy energy is about one order of magnitude larger in hcp cobalt than in the two cubic phases, while the nutation frequency differs by less than a factor of two among them. This preliminary observation suggests that a relation of proportionality may exist between the frequency of nutation and the strength of the magneto-crystalline anisotropy, and that it is sub-linear. We also note that fcc and bcc phases are energetically close; a small change in lattice parameter can induce a so-called Bain transformation between them \cite{cuenya2001observation}. Hence, it is not too surprising that also their magneto-crystalline anisotropy and nutation frequencies are similar. We anticipate that future works will shed light on how to derive the nutation frequency from first principles or from other magnetic properties of the material.

In summary, we measured the temporal evolution of terahertz-field driven spin dynamics in three epitaxial cobalt samples with fcc, bcc and hcp crystal structures. We observed the appearance of THz oscillations with distinct frequencies for the three samples and of a delayed coherent magnetization response, which could be naturally described in the framework of the inertial LLG equation assuming a magnetic damping one order of magnitude larger than the conventional Gilbert damping at FMR frequencies. While surprising, this evidence may be consistent with recent theoretical works suggesting a time-dependent damping coefficient. We could also estimate a constant ratio between the angular momentum relaxation time and the measured damping, in agreement with the prediction of the full relativistic theory of magnetic inertia. Finally, we could observe higher harmonics of the nutation oscillations, not described by the currently accepted inertial LLG equation with temporal derivatives up to the second order, but possibly consistent with a higher order extension of the same equation. Our work provides the strongest evidence for inertial spin dynamics so far, where all the experimental results can be reproduced with a single free parameter. We envisage that our results will trigger future experimental and theoretical investigations towards a deeper microscopic understanding of magnetic inertia at ultrafast time scales.

V.U.  and  S.B.  acknowledge  support  from  the  European Research Council, Starting Grant 715452 “MAGNETIC-SPEED-LIMIT”. R.M. and E.E.F. were supported by U.S. Department of Energy, Office of Science, Office of Basic Energy Sciences, under Contract No. DE-SC0003678.  

\providecommand{\noopsort}[1]{}\providecommand{\singleletter}[1]{#1}%

\end{document}